\documentclass[11pt,a4paper,twoside]{article}

\usepackage{graphicx}
\usepackage{amsmath}
\usepackage{amssymb}

\usepackage[abbr]{harvard}
\citationstyle{dcu}
\citationmode{abbr}

\newcommand{\EXP}[1]{\mathrm{e}^{#1}} 
\newcommand{\DEF}{\overset{\mathrm{def}}{=}}
\newcommand{\DEFt}{\overset{\text{\tiny def}}{=}}
 
\newcommand{\dmat}{\mathrm{d}}

\newcommand{\kete}[1]{|\kern.3ex#1\kern.3ex\rangle}

\newcommand{\brae}[1]{\langle\kern.3ex #1 \kern.3ex|}

\begin{document}

\title{Finding gaps in a spectrum}
\author{Hector Giacomini, Amaury Mouchet\\ \\
Laboratoire de Math\'ematiques 
  et Physique Th\'eorique,\\ Universit\'e Fran\c{c}ois Rabelais de Tours 
--- \textsc{\textsc{cnrs (umr 6083)}},\\
F\'ed\'eration Denis Poisson,\\
 Parc de Grandmont 37200
  Tours,  France. \\{giacomin, mouchet@phys.univ-tours.fr}}

\date{\today}

\maketitle
\begin{abstract}
We propose a method for finding gaps in the spectrum of a 
differential operator. When applied to the one-dimensional  Hamiltonian of the quartic oscillator,
a simple algebraic algorithm is proposed that, step by step, separates with a remarkable precision
all the energies even for a double-well configuration in a tunnelling regime. Our strategy 
may be refined and generalised to a large class of $1d$-problems.
\end{abstract}


\section{Introduction}

Obtaining some physically interesting piece of information on a
spectrum of a differential operator is generically a difficult
problem. For systems in low dimension, say one or two, though
numerical approaches can usually compute the spectrum to a high
precision, very few is usually known when demanding the standard
criterions of rigour of mathematical physics. 

The aim of the present article is to propose a method for finding gaps in the spectrum of a differential operator.
We will mainly focus on the
stationary one-dimensional non-magnetic Schr\"odinger equation with a
smooth potential~$V(x)$ defined on the whole real axis,
\begin{equation}\label{eq:schrodinger1d}
  -\frac{1}{2}\,\frac{\dmat^2\varphi}{\dmat x^2}+V\varphi=E\varphi\;,
\end{equation}
but it can be easily understood that our ideas, which will be
explained in the first part of this article~(\S~\ref{sec:guideline}),
can be extended to a wider family of ordinary differential
equations. For the sake of definiteness, in the second
part~(\S~\ref{sec:quarticoscillator}) we will illustrate more
concretely our method on the quartic oscillator~\eqref{eq:Vq4} where the numerical
``exact'' spectrum will serve us as a touchstone. We will show that we
are able to obtain gaps in the energy spectrum,
\textit{i.e.}~intervals where there is no eigenenergy, with purely
algebraic manipulations involving no integrals. An algorithm is
explicitly constructed in the case of the quartic oscillator --- but
it remains robust under any kind of smooth perturbations --- and provides more and
more non-trivial gaps. As far as we know, in the huge mathematical
physics literature \cite[for instance and their references]{Bender/Orszag78a,Voros83a} 
on the quartic 
oscillator no such gaps have been
found. We will present strong evidence that, for still
unknown reasons, this algorithm allows to resolve all the
eigenenergies with a surprising precision even though, remaining stuck
to the simplest means, we are far from having exhausted all the
possibilities that our method offers.  Actually, some very subtle
information on the spectrum can be gained since we can find a gap in
between the tunnelling doublets, \textit{i.e.}~when the energies are
below the height of the intermediate barrier in a double-well
configuration.

\section{Guideline of the method}\label{sec:guideline}

\subsection{General strategy}\label{subsec:strategy}

The key idea of our method is to construct, for a given integer~$N\geqslant1$, a
real function~$J_N(\varphi',\varphi,x,E)$ from a possible real
square-integrable solution~$\varphi$ of~\eqref{eq:schrodinger1d}
 with energy~$E$
(the prime stands for the derivative~$\dmat/\dmat x$) such that
\begin{eqnarray*}
\mathrm{(i)} &&\frac{\dmat}{\dmat x}\Big(J_N\big(\varphi'(x),\varphi(x),x,E\big)\Big)
               =\big(\varphi(x)\big)^N \, F_{N}(x,E)\;,\\
\mathrm{(ii)} &&\lim_{|x|\to+\infty}J\big(\varphi'(x),\varphi(x),x,E\big)=0.
\end{eqnarray*}
The real function~$F_N$ of both the spatial
coordinate~$x$ and the energy~$E$ is 
$\varphi$\textit{-independent} and is constructed from the
potential~$V$ and its derivatives.  An example of~$J_N$ is given by
equation~\eqref{eq:JN} below and some $F_N$'s are given
by~(\ref{eq:F1} -- \ref{eq:F4}).  Condition~(i) is the cornerstone of our method and,
before justifying how it can be obtained (we will see that condition~(ii) is not so
restrictive), let us first explain how 
gaps in the spectrum may be obtained.

When the conditions~(i) and (ii) are simultaneously
fulfilled, an immediate consequence is that the integral~$\int_{-\infty}^{+\infty}
\big(\varphi(x)\big)^N \, F_N(x,E)\, \dmat x$ vanishes. This implies
that, if $E$ is truly an eigenenergy, the function
$x\mapsto\big(\varphi(x)\big)^N \, F_N(x,E)$ should change its sign.
If~$N$ is even, we obtain a $\varphi$-independent condition:
\textit{for any fixed energy $x\mapsto F_N(x,E)$ must change its sign
on the real axis}.   We will see that for such a
one-dimensional problem, and for a given~$N$, we still can choose~$F_N$
in a wide continuous set of smooth functions on the real axis. A forbidden value of~$E$
(\textit{i.e.} $E$ cannot be an eigenenergy) is
obtained if we are able to construct a~$F_N$ that remains positive on
the whole $x$-axis. Once this property is achieved, it remains stable
 under small perturbations within the set of~$F_N$'s,
and we obtain
a whole interval where no eigenenergy can exist.
 
When~$N$ is odd, \textit{a priori} some non-trivial information can be
extracted for the ground state only, if there is any.  Indeed, it is
known that its wavefunction can be chosen positive for any
Schr\"odinger equation of the form~\eqref{eq:schrodinger1d} (see for
instance \cite[XIII.12]{Reed/Simon78a}) whereas all the excited
eigenfunctions do change their sign. No gap can be found but one can
use this strategy to find upper and lower bounds on the ground state
energy even in a multidimensional 
situation. The differential method presented in \cite{Mouchet05a}
corresponds to~$N=1$.

\subsection{Obtaining condition (i) and determination of~$F_N$}

In order to obtain condition~(i), let us start with~$J$ being a 
homogeneous polynomial of degree~$N$ with respect to its first two
variables~$(\varphi',\varphi)$:
\begin{equation}\label{eq:JN}
J_N\big(\varphi'(x),\varphi(x),x,E\big)=
\sum_{n=0}^N\,a_n(x,E)\,\big(\varphi'(x)\big)^{N-n}\big(\varphi(x)\big)^n\;.
\end{equation}
The smooth functions $\{a_n\}_{n\in\{0,\dots,N\}}$ will be constructed
in order to get condition~(i): Equation~\eqref{eq:schrodinger1d}
allows us to eliminate~$\varphi''$ from the total derivative of~$J_N$
and then, the systematic cancellation of the coefficients of
$(\varphi')^{N-n}$ for~$N-n>0$ leads to the relations
\begin{equation}\label{eq:an}
 \forall n\in\{0,\dots,N-1\},\quad a_{n+1}=-\frac{1}{n+1}\,a'_n-\frac{N-n+1}{n+1}\,2(V-E)\,a_{n-1},
\end{equation}
(to get a unified expression, we define~$a_{-1}\equiv0$) with
\begin{equation}
  F_N(x,E)=a'_N(x,E) + 2\big(V(x)-E\big)\,a_{N-1}(x,E)\;.
\end{equation}
The recurrence relation~\eqref{eq:an} uniquely determines all the
$a_n$'s and $F_N$ from~$a_0$, which remains a free smooth
function. For instance, we have
\begin{align}
 \label{eq:F1}F_1=&  -a''_0+2(V-E)\,a_0\;, \\
 \label{eq:F2}F_2=&\frac{1}{2}\,a'''_0-4(V-E)\,a'_0-2V'a_0\;,\\
 \label{eq:F3}F_3=&-\frac{1}{6}\,a^{(iv)}_0+\frac{10}{3}\,(V-E)\,a''_0+\frac{10}{3}\,V'a'_0
                   +\left(\,V''-6\,(V-E)^2\right)a_0\;,\\
              F_4=&\frac{1}{24}\,a^{(v)}_0-\frac{5}{3}\,(V-E)\,a'''_0 -\frac{5}{2}\,V'a_0''
                   +\left(-\frac{3}{2}\,V''+\frac{32}{3}\,(V-E)^2\right)a'_0\nonumber\\
 \label{eq:F4}    &+\left(-\frac{1}{3}\,V'''+\frac{32}{3}\,V'(V-E)\right)a_0\;.
\end{align}
 The only condition on~$a_0$ is that it must not increase too rapidly
when~$|x|\to+\infty$ in order to get condition~(ii).  From the
standard semiclassical analysis \cite[for instance chap.~VI]{Messiah91a}, we know that if a bound state of~$V$
exists, its wavefunction decreases exponentially as~$\exp(-\big|\int^x
\sqrt{2(V(x')-E)}\, \dmat x'\big|)$ when~$|x|\to+\infty$; therefore the
ansatz~\eqref{eq:JN} will vanish at infinity as soon as $|a_0|$
becomes negligible compared to~$|\varphi|$.  The condition
\begin{equation}\label{eq:a0ii}
  |a_0(x)|\ll \EXP{\left|\int^x \sqrt{2\big(V(x')-E\big)}\, \dmat x'\right|}\quad\mathrm{when}\quad|x|\to+\infty
\end{equation}
is sufficient and not too demanding.

We do not loose 
generality with the form~\eqref{eq:JN}.  In
fact any smooth function $J(\varphi',\varphi,x,E)$ will inevitably
lead to the hierarchy of
functions~$\left\{F_N\right\}_{N\geqslant1}$. To understand this,
consider one monomial $c\,(\varphi')^n\varphi^m$, where $c$ is a function of $x$, in the Taylor
expansion of $\frac{\dmat}{\dmat x} J(\varphi',\varphi,x,E)$ with respect to
$(\varphi',\varphi)$ once $\varphi''$ has been substituted
by~$2(V-E)\varphi$. 
This monomial can be written as
\begin{align}\label{eq:ibp}
  c\,(\varphi')^n\varphi^m=&-\frac{1}{m+1}\left(c\,(\varphi')^{n-1}\right)'\varphi^{m+1}
                         +\frac{1}{m+1}\frac{\dmat}{\dmat x}\left(c\,(\varphi')^{n-1}\varphi^{m+1}\right)\nonumber\\
			 =&-\frac{1}{m+1}\big[c'\,(\varphi')^{n-1}\varphi^{m+1}+2(n-1)(V-E)c\,
                              (\varphi')^{n-2}\varphi^{m+2}\big]\nonumber \\
                         &+\frac{1}{m+1}\frac{\dmat}{\dmat x}\left(c\,(\varphi')^{n-1}\varphi^{m+1}\right).
\end{align}
By using this type of identity an adequate number of times we can
systematically absorb all the powers of $\varphi'$ in a total
derivative while keeping the homogeneity in~$(\varphi',\varphi)$. This
procedure can be pursued until we obtain~$c\,(\varphi')^n\varphi^m$ of
the form $\tilde{c}\,\varphi^{n+m}+\frac{\dmat}{\dmat x}\big(j(\varphi',\varphi,x,E)\big)$
where $\tilde{c}$ and $j$ are smooth functions. A redefinition
of~$J\to J-j$ for each monomial of degree~$n+m=N$ leads to
condition~(i). Rather than the starting point~\eqref{eq:JN}, we could
have started from~$J_N(\varphi',\varphi,x,E)=a_0(\varphi')^N$ and,
working modulo a total derivative, repetitions of procedure~\eqref{eq:ibp}
 would have led to condition~(i).  For small~$N$, it
can been checked that the $F_N$ thus obtained are also given
by~(\ref{eq:F1} -- \ref{eq:F4}).


\section{Application to the quartic oscillator and generalizations}\label{sec:quarticoscillator}

\begin{figure}[!ht]
\center
\includegraphics[width=6.7cm]{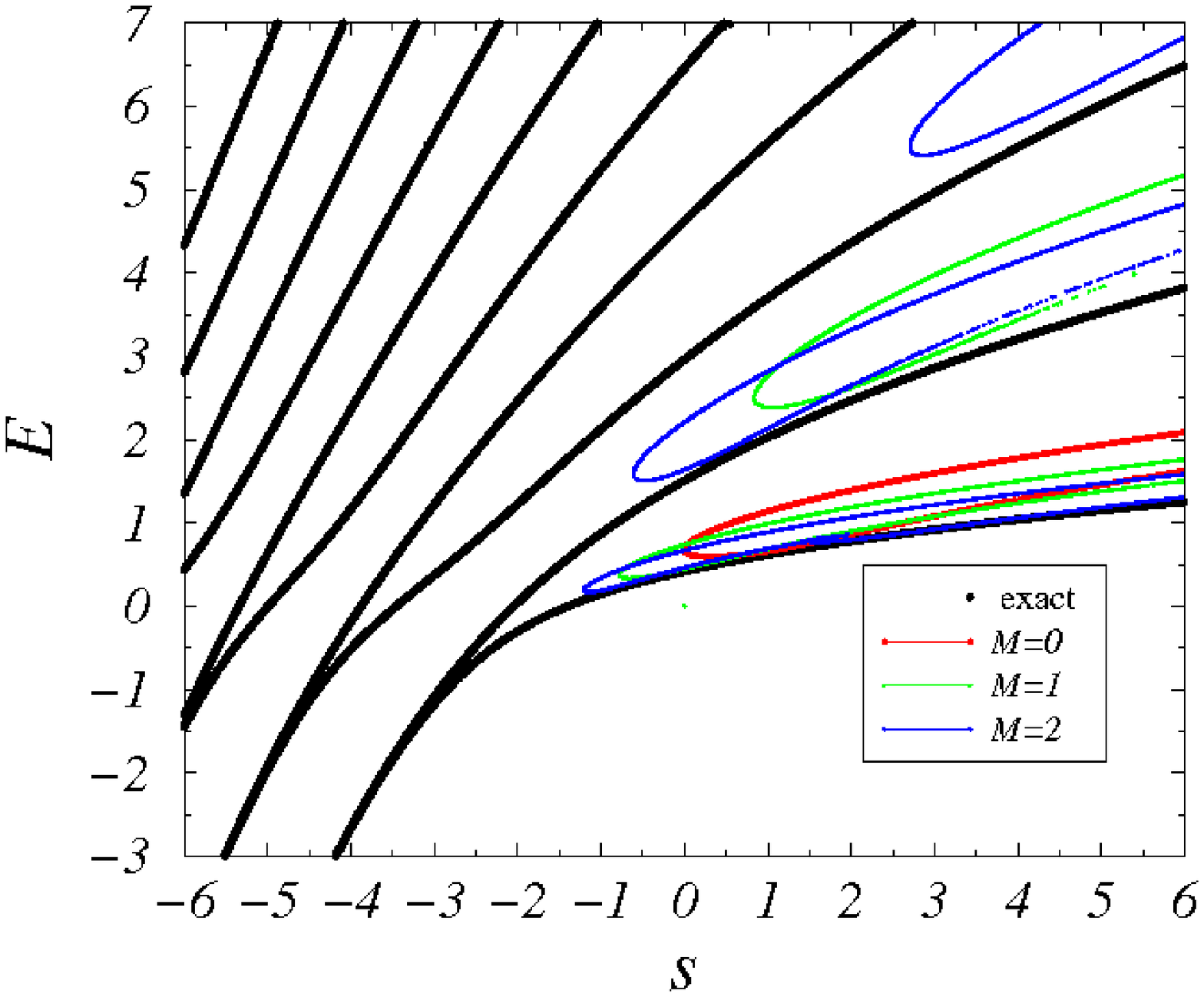}~\includegraphics[width=6.82cm]{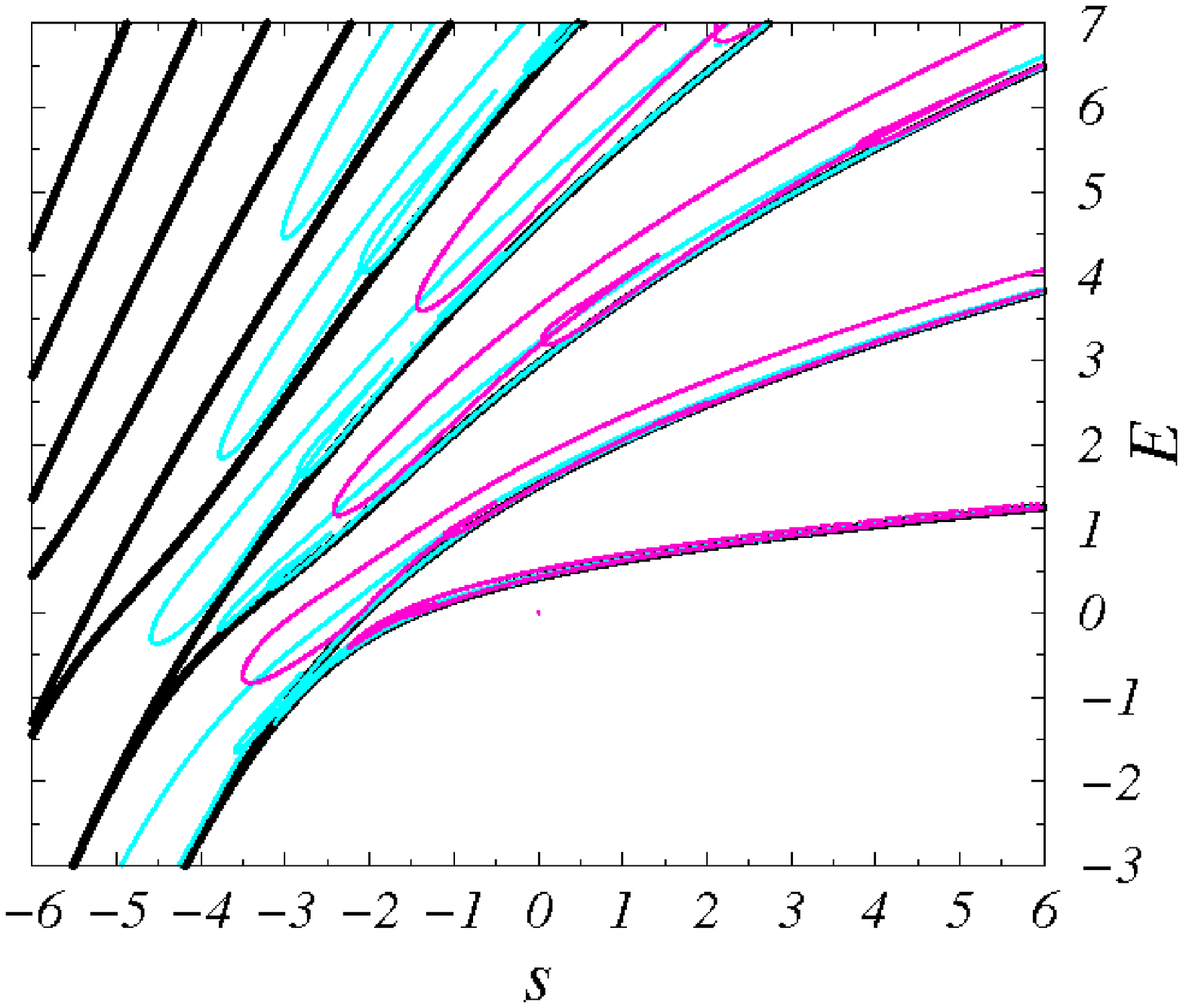}\\\includegraphics[width=6.7cm]{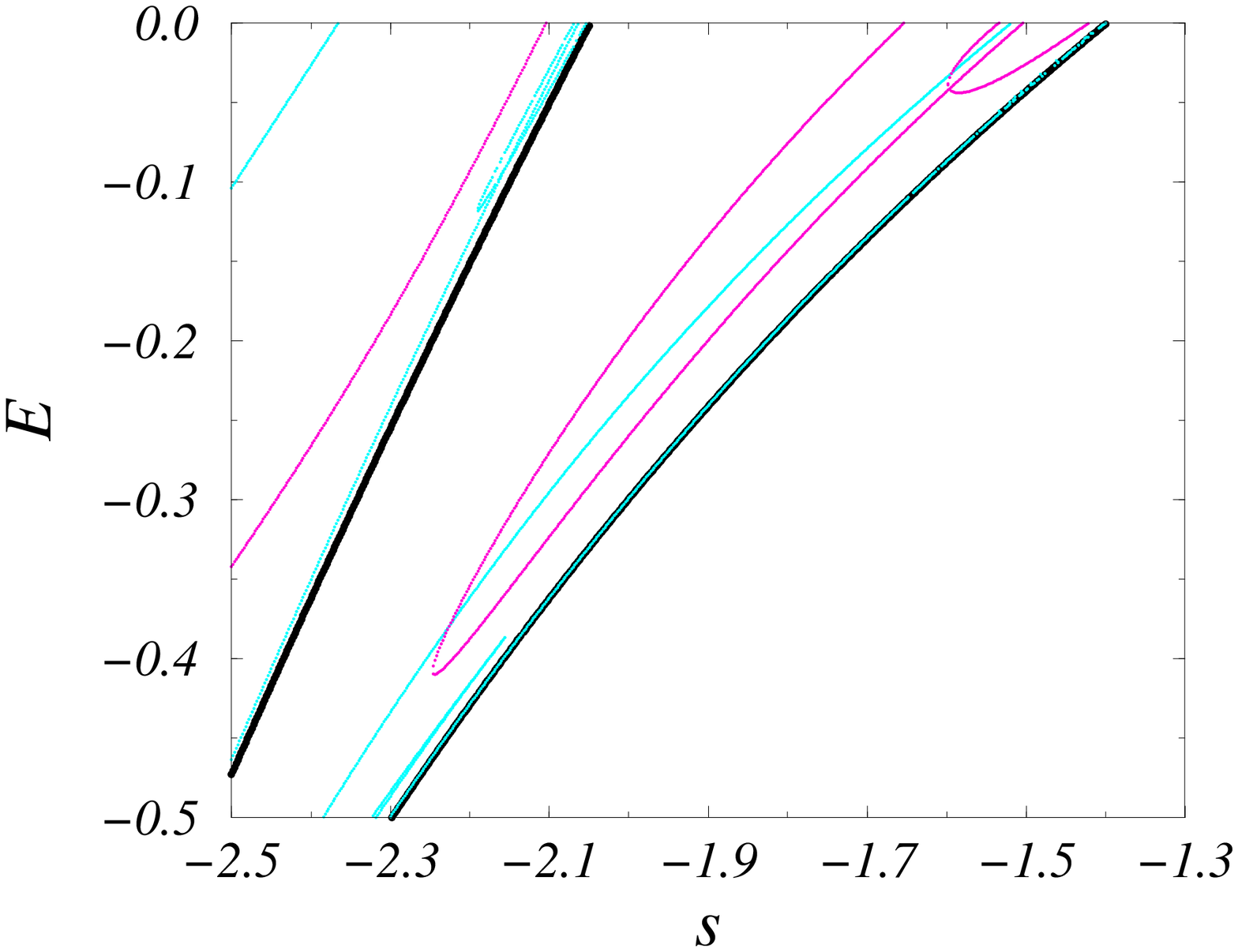}
\caption{\label{fig:gapsM} (color on line) The black thick lines stand
 for the exact spectrum of the Hamiltonian~$p^2/2+sx^2/2+x^4/4$. On
 the upper left pannel the boundaries of the gaps that are given by
 the solutions of \eqref{eq:eqgap} for~$M=0$ (red, the equation of the
 curve is obtained by cancelling \eqref{eq:Delta0}), $M=1$ (green),
 $M=2$ (blue). The gaps on the upper right pannel are obtained for
 $M=7$ (magenta) and $M=15$ (cyan). For a given~$M$, no gap can be
 found for a too small value of~$s$ but we can see that for a given
 value of~$s$ all the lowest energies can be separated by a gap for a
 large enough~$M$.  The boundaries of the gaps converge to the exact
 spectrum when~$M$ increases.  For instance, for~$M=15$ and $s>0$, the
 lowest gap is very thin and can hardly be distinguished from the
 ground state~$E_0$.  In some cases, the equations \eqref{eq:eqgap}
 provide two disconnected gaps in between two successive energies.  As
 can be seen in the pannel below, for~$M=7$ and~$M=15$, even the
 doublet~$(E_0,E_1)$ can be separated in a tunnelling regime
 (\textit{i.e.}  when they are both negative).  }
\end{figure}

To illustrate the efficiency of our approach, in this
section we will consider  a quartic potential, which can always be reduced in appropriate
units to
\begin{equation}\label{eq:Vq4}
  V(x)\ \DEF\  \frac{s}{2}\,x^2+\frac{1}{4}\,x^4,
\end{equation}
for~$s$ real.  The associated energy spectrum is purely discrete and
bounded from below by the minimum $\min(V)$. The energies can been
computed numerically without any difficulty by diagonalizing the
Hamiltonian in the standard basis of the eigenstates of an harmonic
oscillator. They are given by the continuous black lines in
figure~\ref{fig:gapsM}. For large positive~$s$, the bottom of the
spectrum~$\left\{E_n\, |\, n=0,1,2,\dots\right\}$ tends to the
harmonic spectrum~$(n+1/2)\sqrt{s}$. For $s<0$, we get a double-well
configuration whre the central barrier reaches its maximum at~0.  When
$s$ decreases from~$0$, we observe the pairing of the energies
$(E_{2n},E_{2n+1})$ into doublets that characterize tunnelling from
one well to the other. For~$s\leqslant s_0\simeq-2.0481$ the two first
states~$E_0$ and~$E_1$ are both negative, therefore below the
energetic barrier, and~$1/(E_1-E_0)$ represents the tunnelling
oscillation period (recall that we are working in units
where~$\hbar=1$) of a state initially localised in one well and
constructed from a linear combination of the symmetric/antisymmetric
eigenstates associated with~$E_0$ and~$E_1$ respectively. As~$-s$
increases, a standard semiclassical analysis shows that $E_1-E_0$
behaves like~$2^{11/4}|s|^{5/4}\,\exp(-|2s|^{3/2}/3)/\sqrt{\pi}$ for
large~$-s$ \cite[\S~V, for instance]{Garg00a}.  This exponentially
small splitting is a highly non-trivial piece of spectral information
to obtain by approximate methods. Nevertheless, we will show that for
a given positive or negative~$s$, we actually can exhibit a gap
between~$E_0$ and $E_1$.

 For simplicity, and in order to keep the computation tractable with
elementary algebraic manipulations, we will work with~$N=2$ and take
$a_0$ of the form
\begin{equation}\label{eq:a0}
  a_0(x)=P(x)\,\EXP{-\lambda\, x^2/2},
\end{equation}
where $\lambda$ is real (not necessarily positive, see~\eqref{eq:a0ii}) and $P$ is a real polynomial in~$x$ (to lighten
the notations we leave the $E$-dependence
implicit. Note that $P$ may depend on~$\lambda$ as well). Condition~\eqref{eq:a0ii} is largely satisfied. 
From expression~\eqref{eq:F2}, we have~$F_2(x)=Q(x)\exp(-\lambda x^2/2)$ where~$Q$ 
is the polynomial
\begin{multline}\label{eq:Q}
  Q(x)=P'''(x)-3\lambda x P''(x)-\Big(2x^4+(4s-3\lambda^2)x^2+3\lambda-8E\Big)P'(x)\\
   +\Big(2\lambda x^5-(\lambda^3-4s\lambda+4)x^3+(3\lambda^2-8E\lambda-4s)x\Big)P(x),
\end{multline}
that must change its sign if~$E$ is an eigenenergy. To find a forbidden value for the 
energy 
we  construct an even~$Q$ and therefore start from an odd~$P$:
\begin{equation}
  P(x)=\sum_{m=0}^{M} p_m\,x^{2m+1}\;.
\end{equation}  
Then 
\begin{equation}
  Q(x)=\sum_{m=0}^{M+3} q_m\,x^{2m},
\end{equation}
where the coefficients~$\{q_m\}$ are expressed in terms of the~$\{p_m\}$ via the recurrence relation
\begin{multline}\label{eq:qm}
  q_m=(2m+3)(2m+2)(2m+1)\,p_{m+1}
       -(2m+1)\Big(3\lambda(2m+1)-8E\Big)p_m \\
       +\Big(2m(3\lambda^2-4s)-8E\lambda\Big) p_{m-1}
       -(\lambda^3 -4s\lambda+4m-2)\, p_{m-2}
       +2\lambda\, p_{m-3},
\end{multline}
for $m\in\{0,\dots,M+3\}$ (we define~$p_m=0$ for $m<0$ or $m>M$). The simplest way to
control the sign of~$Q$ is to reduce it to 
a polynomial of degree two in $x^2$:
$Q(x)=x^{2M+2}R(x)$ where~$R(x)\DEFt(q_{M+3}\,x^4+q_{M+2}\,x^2+q_{M+1})$. This can be done if we  
 choose the~$p$'s in order to cancel all the $q_m$ for $m\leqslant M$. 
For $m\in\{0,\dots,M-1\}$, equations~\eqref{eq:qm} determine uniquely
 all the $p$'s up to a common factor~$p_0$ than can be taken to one
 without loss of generality: $p_n$ is a polynomial in~$(E,\lambda)$ of
 degree at most~$n$.  From equation~\eqref{eq:qm} written for $m=M$,
 the condition $q_M=0$ imposes an algebraic relation of degree at
 most~$(M+1)$ between~$E$ and~$\lambda$ that implicitly
 defines~$\lambda$ as a function of~$E$ (several branches are possible
 in general). After eliminating~$\lambda$, the discriminant of~$R$
 appears as a function~$\Delta_M(E,s)$ of~$E$ and~$s$ only. For a
 given~$s$, all the values of~$E$ where~$\Delta_M<0$ are
 forbidden. The boundaries of the gaps are given by the zeroes
 of~$E\mapsto\Delta_M(E,s)$. For instance, when $M=0$ we have:
\begin{equation}\label{eq:Delta0}
\Delta_0(E,s)=
65536\,E^6-73728\,E^4s-41472\,E^3+20736\,E^2s^2+7776\,Es+6561.
\end{equation}
For a given value of $s$,
 the real roots of this polynomial define intervals of forbidden values of $E$.
The degree of the polynomial $\Delta_M(E,s)$ increases with $M$.\\
\begin{table}[!ht]
\begin{scriptsize}
\begin{tabular}{|c|c||c||c||c||c||c||c|}
\hline
$n$   & 0             & 1               & 2             & 3             & 4             & 5             & 6            \\
$E_n$ & -0.5012 & -0.2549   & 0.9606  & 2.1003  & 3.5281  & 5.1202  & 6.8609 \\
\hline
\end{tabular}

\bigskip

\begin{tabular}{|l|c|c|c|c|c|c| }
\hline
$M$ & $E_0<\cdots<E_1$ & $E_1<\cdots<E_2$ & $E_2<\cdots<E_3$ & $E_3<\cdots<E_4$ & $E_4<\cdots<E_5$ & $E_5<\cdots<E_6$\\ 
\hline
$\leqslant4$ & $-$ & $-$ & $-$ & $-$ & $-$ & $-$  \\  
\hline
$5$ & $-$ & $[0.0399,0.4582]$ & $-$ & $-$ & $-$ & $-$  \\  
\hline
$6$ & $-$ & $[-0.1108,0.4359]$ & $-$ & $-$ & $-$ & $-$ \\
\hline
$7$ & $-$ & $[-0.1836,0.3908]$ & $[1.1734,1.4748]$ & $-$ & $-$ & $-$ \\
\hline
$8$ & $[-0.4614,-0.4281]$ & $[-0.2001,0.3408]$ & $[1.0601,1.5156]$ & $-$ & $-$ & $-$ \\
\hline
$15$ & $[-0.4994,-0.4865]$ & $[-0.2414,0.0507]$ & $[0.9845,1.2930]$ & $[2.1214,2.3143]$ & $[3.7509,4.3535]$ & $[5.3268,6.0726]$ \\
 & $[-0.4835,-0.4336]$ &  &  & $[2.3789,2.7014]$ &  &  \\
\hline
$16$ & $[-0.5006,-0.4369]$ & $[-0.247,-0.24]$  & $[0.9903,1.2652]$ & $[2.1318,2.3167]$ & $[3.6068,3.7792]$& [5.3426,6.0835] \\
 &  & $[-0.2325,0.0206]$ & &  $[2.3307,2.6590]$ &$[3.9847,4.3128]$&  \\
\hline
$\displaystyle\bigcup_{M=0}^{16}$  & $[-0.5006,-0.4211]$ & $[-0.2497,0.4582]$  & $[0.9726,1.5156]$ & $[2.1214,2.8710]$ & $[3.6068,4.4028]$& [5.3268,6.0835] \\
\hline
\end{tabular}
\end{scriptsize}

\caption{\label{tab:gaps} Comparison between the exact
energies~$\left\{E_n\, |\, n=0,1,2,\dots\right\}$ (given in the upper
table) and the gaps for~$s=-2.3$ and several values of~$M$.  The two
lowest energies~$(E_0,E_1)$, being negative though below the central barrier,
 form a tunnelling
doublet. There are two gaps between $(E_0,E_1)$ for~$M=15$, and one
gap for~$M=16$. The last line sums up the best bounds when comparing
the gaps obtained for each~$M$ up to~$16$. We did not retain the lowest
 bounds that are below the trivial bound~$\min V$.
  }
\end{table}

We can generalize and reformulate these algebraic manipulations to any
kind of polynomial potential of even degree~$\deg V$ (not necessarily
symmetric).  With the choice~\eqref{eq:a0}, the
expression~\eqref{eq:F2} shows that $F_2\,\EXP{\lambda\, x^2/2}$ is a
polynomial~$Q(x,\lambda,E)$ of degree~$\deg V+\deg P+1$ in~$x$. For
any $(\lambda,E)$ the coefficients of~$P$ can be chosen to cancel all
the coefficients of~$x\mapsto Q(x,\lambda,E)$ but the $(\deg V+2)$th
higher powers. The cancellation of the coefficient~$q(\lambda,E)$ of
$x^{\deg P}$ allows the factorization~$Q(x,\lambda,E)=x^{\deg
P+1}R(x,\lambda,E)$ where the degree of~$x\mapsto R(x,\lambda,E)$ is
precisely~$\deg V$. Now, if~$\deg P$ is odd and $x\mapsto
R(x,\lambda,E)$ has no root of odd multiplicity for a given
couple~$(\lambda,E)$ such that~$q(\lambda,E)=0$, we are sure that~$E$
cannot be an eigenvalue.  In the $(x,\lambda,E)$-space, the two
equations~$q(\lambda,E)=0,R(x,\lambda,E)=0$ define generically a
curve~$\mathcal{C}$ (possibly made of disjoint smooth pieces) that
can be parameterized by~$x$, namely $\big(E(x),\lambda(x)\big)$, where
the Jacobian~$J=|\partial_\lambda q\,\partial_ER-\partial_\lambda
R\,\partial_Eq|$ does not vanish. The projection of~$\mathcal{C}$ on
the $E$-axis defines some intervals outside which no eigenenergy can
be found.  The borders of these intervals are the projections of some
points (not necessarily of all points) of~$\mathcal{C}$ where the
tangent is normal to the~$E$-axis. Using the implicit function theorem
these points are to be found among the solutions
of~$0=dE/dx=-\partial_\lambda q\,\partial_xR/J$.  To sum up, the
boundaries of the gaps are to be found from the solutions --- if there
are any --- of the  three equations
\begin{equation}\label{eq:eqgap}
  q(\lambda,E)=0,\quad R(x,\lambda,E)=0,\quad\partial_\lambda q(\lambda,E)\,\partial_x R(x,\lambda,E)=0.
\end{equation}
The boundaries obtained above for the quartic potential using the
zeroes of the discriminants~$\Delta_M$ are included in the solutions
of~\eqref{eq:eqgap} since they cancel simultaneously $q$, $R$
and~$\partial_x R$ but some others boundaries may be obtained
if~$\partial_\lambda q$ vanishes instead of~$\partial_x R$.
Determining which values represent boundaries of gaps from all the
solutions of \eqref{eq:eqgap} may require some global analysis that
can be pursued numerically.  The larger~$M$, the larger the degree of
the algebraic equations to solve and the more gaps are expected to be
found. For the quartic oscillator, the numerical results are shown for
several values of~$M$ in figure~\ref{fig:gapsM} and
table~\ref{tab:gaps} illustrates the precision of our method.

Let us consider also the even potential~$V(x)=x^6$. The
polynomial $R$ is now of third degree in the variable $x^2$. As explained
above, a necessary condition to obtain the boundaries of the gaps for
a given positive integer $M$ is to solve simultaneously the three
equations \eqref{eq:eqgap}.  The last step consists in verifying that
the polynomial $R(x)$ does not have real roots of odd
multiplicity. The results obtained are qualitatively similar that the
one's showed above for the quartic oscillator (see
tables~\ref{tab:gapsx6}).

\begin{table}[!ht]
\begin{scriptsize}
\begin{tabular}{|c|c||c||c|}
\hline
$n$   & 0             & 1               & 2               \\
$E_n$ & 0.6807 &  2.5797  & 5.3948  \\
\hline
\end{tabular}\hfill
\begin{tabular}{|l|c|c|}
\hline
$M$ & $E_0<\cdots<E_1$ & $E_1<\cdots<E_2$ \\ 
\hline
$0$ & $-$ & $-$  \\  
\hline
$1$ & $[0.967,1.041]$ & $-$  \\  
\hline
$2$ & $[0.736;1.122]$ & $-$ \\
\hline
$3$ & $[0.710,1.081]$ & $[3.016,3.624]$  \\
\hline
$4$ & $[0.715,1.036]$ & $[2.724,3.702]$  \\
\hline
$5$ & $[0.723,0.995]$ & $[2.674,3.642]$  \\
\hline
\end{tabular}
\end{scriptsize}
\caption{\label{tab:gapsx6} Comparison between the exact 
energies~$\left\{E_n\, |\, n=0,1,2\right\}$ (given in the left table) and
 the gaps for~$V(x)=x^6$.
 }
\end{table}

\section{Conclusions}\label{sec:conclusion}

From what precedes it is clear that our method is not specific to the
quartic potential but can be adapted straightforwardly to many other
situations.

This method can be also applied to a spectral problem with different
boundary conditions. For, say, Dirichlet boundary conditions
equation~(ii) is replaced by the vanishing of~$J_N$ at the two points
where the function~$\varphi$ vanishes.

Even though our method cannot provide a rigorous proof of the
existence of an eigenenergy in a given range, it offers some precise
clues where some possible energies may lie, specially if some
convergence behaviour is observed, as it appears in the case of the
quartic oscillator. Therefore we are able to obtain this way a substantial piece of physical
information.

\textbf{Aknowlegments:}
We would like to thank J\'er\'emy Ledeunff for helping us to check the eigenenergies for the~$x^6$-potential. 



\end{document}